\documentclass[11pt]{article}
\usepackage{graphicx}

\newsavebox{\hflrar}
\sbox{\hflrar}{\makebox[0pt][l]
{${\scriptstyle \leftharpoonup}$}{${\scriptstyle \rightharpoonup}$}}

\def \to {\rightarrow}

\begin{document}
\pagestyle{plain}
\vskip 10mm
\begin{center}
{\bf\Large NRQCD Factorization and Universality of NRQCD Matrix Elements } \\
\vskip 10mm
J.P. Ma   \\
{\small {\it Institute of Theoretical Physics , Academia
Sinica, Beijing 100080, China \ \ \ }} \\
{\small {\it Department of Physics, Shandong University, Jinan Shandong 250100, China}}
\\
Z.G. Si \\
{\small {\it Department of Physics, Shandong University, Jinan Shandong 250100, China}}
\end{center}
\vskip 0.4 cm
%%%%%%%%%%%%%%%%%% abstract of this paper %%%%%%%%%%%%%%%%%%%%%%%%%%%%%%%%

\begin{abstract}
The approach of nonrelativistic QCD(NRQCD) factorization was proposed to study inclusive
production of a quarkonium.
It is widely used and successful.
However, a recent study
of gluon fragmentation into a quarkonium at two-loop level shows
that the factorization is broken.
It is suggested that the color-octet NRQCD matrix elements
should be modified by adding a gauge link to restore the factorization.
The modified matrix elements may have extra soft-divergences at one-loop level which
the unmodified can not have, and this can lead to a violation of
the universality of these matrix elements.
In this letter, we examine in detail the  NRQCD factorization for inclusive quarkonium production
in $e^+ e^-$ annihilation at one-loop level. Our results
show that the factorization can be made without the modification of NRQCD matrix elements
and it can also be made for relativistic corrections. It turns out that the suggested gauge
link will not lead to nonzero contributions to color-octet NRQCD matrix elements at one-loop level
and at any order of $v$.
Therefore the universality holds at least at one-loop level.

\end{abstract}
\par\vfil\eject

\par
A quarkonium system provide a unique place
to study the dynamics of QCD, because a quarkonium mainly consists of a heavy quark pair $Q\bar Q$
and they move with a small velocity $v$. An extensive review of quarkonium physics can be found
in \cite{QWG}.
A decade ago
the approach of NRQCD factorization was proposed to study  inclusive production of a quarkonium\cite{NRQCD}.
In this approach the production of a $Q\bar Q$ pair can be studied with perturbative QCD because
the mass $m$ of $Q$ provides a large scale, the formation
of the $Q\bar Q$ pair into a quarkonium is characterized with NRQCD matrix elements by an expansion
in $v$.
In order to have prediction power these matrix elements should be universal,
i.e., they do not depend on how the $Q\bar Q$ pair is produced.
This leads to the NRQCD factorization. This approach is widely used and successful.
Especially, it can systematically take higher Fock state components of a quarkonium, including
those components which contain a heavy quark pair in color-octet, into account.
A striking success of the approach is to explain the $\psi'$ anomaly at Tevatron \cite{CrosEx} by taking
color-octet components into account\cite{BrFl}.
\par
Although the approach is successful, a complete proof
of the factorization does not exist. Studies of various processes at one-loop level
really show that the factorization hold at one-loop level\cite{Ma1,BraCh,BraLe,NLOP,BMR,KK}.
But a recent study shows that the factorization for gluon fragmentation into a quarkonium
at two-loop level is incomplete, indicated by that some uncancelled infrared(I.R.) divergences
are not matched by NRQCD matrix elements.
To restore the factorization,
a gauge link is introduced to modify color-octet NRQCD matrix elements for matching these I.R. divergences\cite{Qiu}.
At one-loop level, this gauge link can generate some extra I.R. divergences in the modified
matrix elements than those unmodified. This can affect the factorization at one-loop level in the cases
studied before and can lead to a violation of the universality of color-octet NRQCD matrix elements.
\par
It is the purpose of the letter to examine if the universality is lost
and if the factorization can be done  in inclusive production
of a quarkonium through $e^+e^-$-annihilation through a color-octet $Q\bar Q$ pair.
Our analysis includes not only the leading contributions in the $v$-expansion, but also
the contribution from relativistic corrections at order of $v^2$.
We show at one-loop
level in detail how soft divergences are cancelled or matched by color-octet NRQCD matrix elements
without the gauge link suggested in \cite{Qiu}.
Our study also shows that the relativist correction for a color-octet $Q\bar Q$ pair can also
be factorized in the same way. To our knowledge, there
is no known example studied at one-loop level to show that the NRQCD factorization holds.
\par
We consider the inclusive production of a quarkonium $H$ in the process
\begin{equation}
e^+ e^- \to \gamma^* \to H +X,
\end{equation}
where the virtual photon is with the momentum $q_\gamma$ and $q^2_\gamma$ is much larger
than the square of the quarkonium mass.
For this process we need to calculate the tensor
\begin{equation}
  T^{\mu\nu}(P,q_\gamma,H) = \sum_X \int d^4 x e^{iq_\gamma \cdot x} \langle 0 \vert J^\nu (0) \vert H(P) +X \rangle
                      \langle H (P) + X \vert J^\mu (x) \vert 0 \rangle
\end{equation}
where the quarkonium carries $P$ and $J^\mu$ is the electric current. The NRQCD factorization
in \cite{NRQCD} suggests the tensor can be written in a factorized form:
\begin{eqnarray}
T^{\mu\nu} (P,q_\gamma,H)  &=&   F^{\mu\nu}(^1S_0)\cdot \langle 0 \vert O_8 (^1 S_0, H) \vert 0 \rangle
       + G^{\mu\nu}(^1S_0) \cdot \langle 0 \vert P_8 (^1 S_0, H) \vert 0 \rangle
\nonumber\\
   && + F^{\mu\nu} (^3P_0) \cdot \langle 0 \vert O_8 (^3 P_0, H ) \vert 0 \rangle
   + F^{\mu\nu,ijkl} (^3P_2)\cdot \langle 0 \vert O_8^{\{ij\},\{kl\}} (^3 P_2, H ) \vert 0 \rangle
   +\cdots,
\end{eqnarray}
where the matrix elements in the right hand side are defined  with NRQCD fields and can be found
in \cite{NRQCD}. We only consider the production through
those color-octet channels which can be at the leading order of $\alpha_s$,
i.e., the channel with the quantum number $^1S_0$ and $^3P_J$ with $J=0,2$.
There is a velocity-power counting rule to determine the relative importance
of NRQCD matrix elements\cite{pv,NRQCD} for a given quarkonium.
In the factorized form the second term is for the relativistic correction of the
channel $^1S_0$.
The coefficients in the front of the NRQCD matrix elements can be calculated with perturbative
QCD, they are series in $\alpha_s$, e.g.,
\begin{equation}
G^{\mu\nu}(^1S_0)=G_0^{\mu\nu}(^1S_0) + G_1^{\mu\nu}(^1S_0) + \cdots,
\end{equation}
where the subscriber $0(1)$ stand for the tree(one-loop) contribution.
If the factorization holds, these perturbative coefficients should not contain any I.R. divergence.
To determine these coefficients,
one replaces the quarkonium $H$ with a $Q\bar Q$ state and calculates $T^{\mu\nu}$ and
the NRQCD matrix elements. By comparing both sides of Eq.(3) calculated with the $Q\bar Q$ pair
the perturbative coefficients can be extracted.
If the factorization holds, soft divergences in $T^{\mu\nu}$ will have the same
form as those appearing in the matrix elements so that the perturbative coefficients do not
contain any soft divergence.
\par
To study the factorization we need to calculate the tensor $T^{\mu\nu}$ in perturbative theory after replacing
the quarkonium $H$ with those $Q\bar Q$ states:
\begin{equation}
  T^{\mu\nu} (P,q_\gamma,Q\bar Q) =
   \sum_X \int d^4 x e^{iq_\gamma\cdot x} \langle 0 \vert J^\nu (0) \vert Q(p_1')\bar Q(p_2') +X \rangle
                      \langle \bar Q(p_2) Q(p_1) + X \vert J^\mu (x) \vert 0 \rangle.
\end{equation}
It should be noted that the heavy quarks carry different momenta
in the amplitude and its complex conjugated. This will enable us to identify different
states of the $Q\bar Q$ pair. These momenta
are given as:
\begin{equation}
p_1 =\frac{1}{2} P + q, \ \ \ \ \ \  p_2 = \frac{1}{2} P -q, \ \ \ \ \ \
p_1' =\frac{1}{2} P + q', \ \ \ \ \ \  p_2' = \frac{1}{2} P -q'.
\end{equation}
In the rest frame of the $Q\bar Q$, $q^\mu = (0, {\bf q})$ and $q^{\prime \mu} = (0, {\bf q}')$ with
${\bf q}^2 ={\bf q}^{\prime 2}$ and
\begin{equation}
   P^2= (p_1+p_2)^2 =4 (m^2 + {\bf q}^2 ), \ \ \ \ \ \   {\bf v} =\frac{{\bf q}}{m}, \ \ \ \  {\bf v'}=\frac{{\bf q^\prime}}{m},
\end{equation}
$v(v')$ is the velocity of the heavy quark in the rest frame.
At tree level, the unobserved state $X$ contains only one gluon. By expanding ${\bf v}$ and ${\bf v'}$
and identifying quantum numbers, one can determine  four coefficients in Eq.(3), i.e.,
\begin{eqnarray}
T^{\mu\nu}_0(P,q_\gamma, Q\bar Q) & =&  F_0^{\mu\nu}(^1S_0)\cdot \langle 0 \vert O_8 (^1 S_0, Q\bar Q) \vert 0 \rangle
   +  G_0^{\mu\nu}(^1S_0) \cdot \langle 0 \vert P_8 (^1 S_0, Q\bar Q) \vert 0 \rangle
\nonumber\\
  &&  + F_0^{\mu\nu} (^3P_0) \cdot\langle 0 \vert O_8 (^3 P_0, Q\bar Q) \vert 0 \rangle
   +  F_0^{\mu\nu,ijkl} (^3P_2)\cdot \langle 0 \vert O_8^{\{ij\},\{kl\}} (^3 P_2, Q\bar Q) \vert 0 \rangle
\nonumber\\
   && +{\mathcal O}(v^4).
\end{eqnarray}
In the expansion in ${\bf v}$ and ${\bf v'}$, the leading terms give contributions to
$F_0^{\mu\nu}(^1S_0)$ for the $^1S_0$ state, the next-to-leading terms, which are linear
in ${\bf v}$ and ${\bf v'}$ like ${\bf v}\cdot{\bf v'}$, give contributions to $F_0^{\mu\nu} (^3P_0)$
for the $^3P_0$ state
and $F_0^{\mu\nu,ijkl} (^3P_2)$ for the $^3P_2$ state.
The next-to-next-to-leading terms are proportional either to the tensor $v^i v^j$ or
to $v^{\prime i}v^{\prime j}$. One can decompose the tensor $v^i v^j$ or
to $v^{\prime i}v^{\prime j}$ into the component of the $S$-wave with $l=0$ and
of the $D$-wave with $l=2$. The $S$-wave component corresponds to the
relativistic correction of the $^1S_0$ state.
At tree level all these coefficients are nonzero and contain no soft divergence, their detailed
forms are not important for our purpose because we will show
that the soft-divergent correction at one-loop to these coefficients is
proportional to the tree-level result $T^{\mu\nu}_0$.
We will use Feynman gauge in this letter. In this gauge
one can clearly see how soft divergences are cancelled or matched in a diagram-by-diagram manner.
\par
One-loop corrections consist of two parts. One is the virtual correction, another
is the real correction in which the unobserved state $X$ contains two gluons or a light
quark pair. Beside corrections from wave-function renormalization there are many Feynman diagrams.
Since we are only interested in soft divergences, we do not need to consider all diagrams but only those
containing soft divergences. The diagrams with soft divergences are given in Fig.1..
To obtain the soft-divergent parts of these diagrams we employ the eikonal approximation with some
modification.
\par
%%%%%%%%%%%%%%%% Inset Fig. 1 here %%%%%%%%%%%%%%%%%%%%%%%%%%%%%%

\begin{figure}[hbt]
\begin{center}
\includegraphics[width=9cm]{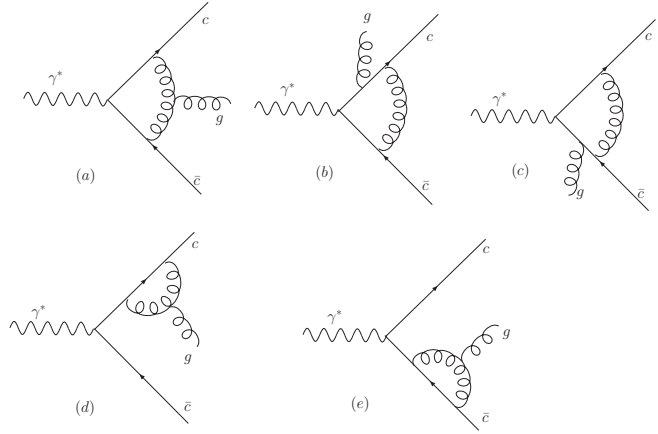}
\end{center}
\caption{ Diagrams for contributions containing soft divergences.}
\label{Feynman-dg2}
\end{figure}
%\end{center}
\par
To illustrate the eikonal approximation used here, we consider
the contribution from Fig.1a to the matrix element
$R^\mu = \langle \bar Q(p_2) Q(p_1),G(k,\varepsilon^*, a) \vert J^\mu (x) \vert 0 \rangle$:
\begin{eqnarray}
R^\mu_{1a} &=&  \int \frac{ d^4 k_1}{(2\pi)^4 } \bar u(p_1) ( -i g_s T^b \gamma ^{\rho})
     \frac{\gamma\cdot(p_1+k_1)+m}{(p_1+k_1)^2-m^2} (-ie Q \gamma^\mu)
     \frac{\gamma \cdot (-k_2 -p_2)+m}{(p_2+k_2)^2 -m^2} (-ig_s T^c \gamma^\sigma ) v (p_2)
\nonumber\\
   && \cdot \frac{1}{k_1^2} \frac{1}{k_2^2}(-g_s f^{abc} ) \left [
    (-k-k_1)_\sigma \varepsilon^*_\rho +(k_1-k_2)\cdot \varepsilon^* g_{\rho\sigma}
    +(k_2+k)_\rho \varepsilon^*_\sigma \right ] ,
\end{eqnarray}
with $k_2 =k- k_1$. The soft divergence appears when $k_1$ or $k_2$ becomes soft, i.e.,
all components of $k_1$ or $k_2$ becomes small, and when $k_1$, hence also $k_2$, is collinear
to the momentum $k$ of the outgoing gluon.
If the gluon with $k_1$ is soft, the standard approximation is to neglect all $k_1$ in nominators and keep
only the leading order in $k_1$ in the denominators.
Therefore the contribution from the soft region of $k_1$ can be written as:
\begin{eqnarray}
R^\mu_{1a,s1}  &=& -i eQ g_s^3 \int \frac{ d^4 k_1}{(2\pi)^4 }
 \bar u(p_1) f^{abc} T^b T^c \frac{2 p_1^\rho}{2 p_1\cdot k_1} \gamma^\mu \frac{\gamma\cdot(-p_2 -k)+m}{2 p_2 \cdot k}
 \gamma^\sigma v(p_2)
\nonumber\\
 && \cdot \frac{1}{k_1^2}\cdot \frac{1}{(-2 k_1\cdot k)}\cdot \left [ -k_{\sigma} \varepsilon_\rho^*
     -k \cdot \varepsilon^* g_{\rho\sigma}+2 k_\rho \varepsilon^* _\sigma \right ].
\end{eqnarray}
We use dimensional regularization with $d=4-\epsilon$ to regularize divergences.
The pole with $1/\epsilon_{I}$ is for I.R. divergence with $\epsilon_I=4-d$.
Other poles without the subscriber $I$
are for U.V. divergences. Calculating the integral we find
that the soft amplitude contains double pole in the form $\epsilon^{-1} (\epsilon^{-1} -\epsilon^{-1}_I)$,
indicating that the approximation may not be convenient here. A convenient approximation
we will use is to keep denominators exact. With this approximation,
the soft- and collinear-divergent part of the contribution can be written as
\begin{eqnarray}
R^\mu_{1a,sc} &=& -i eQ g_s^3 \int \frac{ d^4 k_1}{(2\pi)^4 } \frac{1}{(2p_1\cdot k_1 +i0)(2p_2\cdot k_2 +i0)
                           (k_1^2 +i0) (k_2^2 +i0)}
\nonumber\\
   &&  \bar u(p_1) f^{abc} T^b T^c
       \left [  (2p_1^\rho +\gamma^\rho \gamma^+ k_1 ^-) \gamma^\mu
        ( -2p_2^\sigma -k_2^- \gamma^+ \gamma^\sigma ) \right ]   v (p_2)
\nonumber\\
  && \left [ (-k-k_1)^- n_\sigma \varepsilon^*_\rho +(2k_1 -k)^- n\cdot \varepsilon^* g_{\rho\sigma}
    +(k_2+k)^- n_\rho \varepsilon^*_\sigma \right ] ,
\end{eqnarray}
and $R^\mu_{1a} -R^\mu_{1a,sc}$ contains no
any soft divergence. In the above equation we have taken a frame in which
\begin{equation}
k^\mu =(0,k^-,0,0)=k^- n^\mu, \ \ \ \ \ \ q^\mu_\gamma =(q_\gamma^+, q^-_\gamma, 0,0).
\end{equation}
It should be noted that $R^\mu_{2a,sc}$ can be written in a covariant form.
Performing a similar analysis for other diagrams and loop-momentum integration we obtain
the soft-divergent part of the one-loop correction to $R^\mu$.
The sum of contributions from Fig.1a, Fig.1d and Fig.1e can be written in a compact
form:
\begin{eqnarray}
R^\mu_{1a,sc}+R^\mu_{1d,sc}+R^\mu_{1e,sc} &=& \frac {i eQ g_s^3}{(4\pi)^2} N_c
\left \{
2 \left ( \frac{2}{\epsilon_{I}}\right )^2  -\frac{2}{\epsilon_{I}} \left [ 2\gamma -2
 + \ln\frac{(2p_1\cdot k)^2}{4\pi m^2 \mu^2 } + \ln\frac{(2p_2\cdot k)^2}{4\pi m^2 \mu^2 } \right ]\right\}
\nonumber\\
&& \cdot \bar u(p_1) T^a \left [ \gamma^\nu \frac{\gamma\cdot(p_1+k)+m}{(p_1+k)^2 -m^2}
\gamma^\mu + \gamma^\mu \frac{\gamma\cdot(-p_2 -k)+m}{(p_2+k)^2 -m^2} \gamma^\nu \right ] v(p_2),
\end{eqnarray}
i.e., the soft-divergent part is proportional to the tree-level amplitude.
The double pole in $\epsilon_I$ is from the overlap of collinear- and soft region
of the loop momentum.
The contribution from Fig.1b and Fig.1c contain not only soft divergences but also
Coulomb singularities. With our approximation the sum of these two diagrams gives:
\begin{eqnarray}
R^\mu_{2b,cs} +R^\mu_{2c,cs} &=& \frac{i eQ g_s^3}{(4\pi)^2} \frac{1+2 v^2}{N_c}
\left [ \frac{\pi^2}{2 v}
               + \frac{2}{\epsilon_{I}} \left ( 1-\frac{2v^2}{3} \right )  \right ]
\nonumber\\
&& \cdot
  \varepsilon^*_\nu \bar u(p_1) T^a \left [ \gamma^\nu \frac{\gamma\cdot(p_1+k)+m}{(p_1+k)^2 -m^2}
\gamma^\mu + \gamma^\mu \frac{\gamma\cdot(-p_2 -k)+m}{(p_2+k)^2 -m^2} \gamma^\nu \right ] v(p_2) +\cdots,
\end{eqnarray}
where $\cdots$ stand for terms which are finite with $\epsilon, v \to 0$, and higher orders in $v$.
\par
Putting everything together, we obtain the soft divergent part of the
virtual one-loop contribution to $T^{\mu\nu}$ in Feynman gauge:
\begin{eqnarray}
T_{1,\ vir.}^{\mu\nu}(P,q_\gamma, Q\bar Q) &=& -\frac{\alpha_s}{2\pi}
  \frac{1}{N_c}\left [  \frac{2}{\epsilon_{I}} \left ( 1+\frac{2}{3} v^2 +\frac{2}{3} v'^2 +{\mathcal O}(v^4 ) \right )+
  \frac{\pi^2}{2 v} (1+ {\mathcal O}(v^2 ) ) \right ] T_0^{\mu\nu}(P,q_\gamma, Q\bar Q)
\nonumber\\
 &&   -\frac{\alpha_s}{2\pi} N_c \left [
 \left ( \frac{2}{\epsilon_{I}}\right )^2  -\frac{2}{\epsilon_{I}} \left ( \gamma -1
 + L_n
 \right ) \right ] T_0^{\mu\nu}(P,q_\gamma, Q\bar Q)
\nonumber\\
   && -\frac{\alpha_s}{2\pi}
  \left ( \frac{2}{\epsilon_I} \right )\left [ \frac{N_c^2-1}{N_c} + \left ( \frac{5}{6} N_c
   - \frac{1}{3} N_f \right ) \right ]
   T_0^{\mu\nu}(P,q_\gamma, Q\bar Q) +\cdots
\nonumber\\
   L_n &=& \frac{1}{2}\ln\frac{ (p_1\cdot k)(p_1'\cdot k)(p_2\cdot k)(p_2'\cdot k)}{(\pi m^2\mu^2)^2 } ,
\end{eqnarray}
where $\cdots$ denote  finite parts and
the correction terms in the third line come from wave-function renormailization of the heavy quarks
and the gluon.
\par
Now we turn to the real correction. The real correction consists of amplitudes
of two gluons or a light quark pair. The two gluons can be emitted by heavy quarks and by gluon splitting,
the light quark pair can only be generated through gluon splitting. It should be noted that
a light quark pair can also be produced in other ways, but the contribution of this
is irrelevant for the factorization discussed here, because
in this case the heavy quark pair is in states other than those given in Eq.(8) and the contribution
at order $\alpha_s^2$ is free from soft divergences.
The Feynman diagrams for the
real correction is given in Fig.2.
\par
%%%%%%%%%%%%%%%% Inset Fig. 1 here %%%%%%%%%%%%%%%%%%%%%%%%%%%%%%

\begin{figure}[hbt]
\begin{center}
\includegraphics[width=9cm]{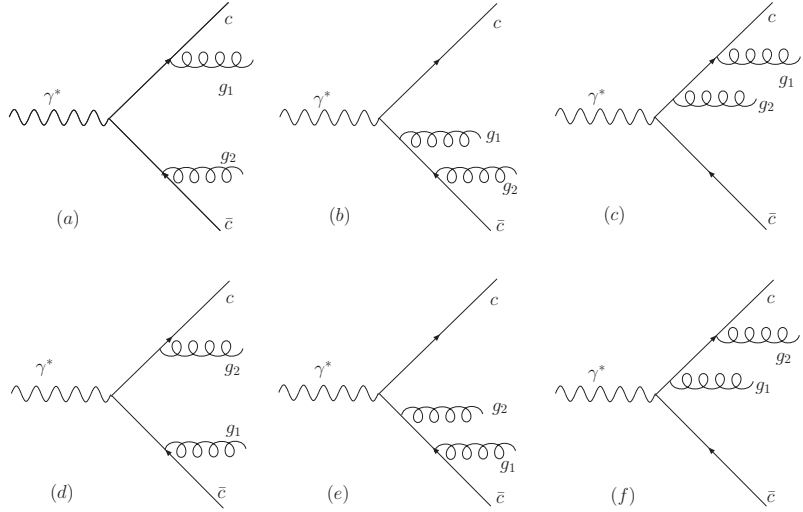}
\end{center}
\caption{ Real correction}
\label{Feynman-dg3}
\end{figure}
%\end{center}
\par
We write the amplitude with two gluons in the final state as
\begin{equation}
R^\mu_{gg} = R^\mu_{gg,Q} + R^\mu_{gg,g},
\end{equation}
where the first term is only for two gluons emitted by heavy quarks and second is for gluon splitting.
If two gluons are emitted by heavy quarks, there are only I.R. divergences, when
one of the gluons becomes soft. Collinear singularities will not appear here because
of the massive quarks. When the gluon with the momentum $k_1$ becomes soft, the soft contribution
of $R^\mu_{gg,Q}$ with the standard eikonal approximation can be written:
\begin{equation}
R^\mu_{gg,Q,s}= \frac{i g_s }{2} f^{a_1a_2 a} \left ( \frac{p_1\cdot\varepsilon^*_1}{p_1\cdot k_1}
 + \frac{p_2\cdot\varepsilon^*_1}{p_2\cdot k_1} \right )
 \langle \bar Q(p_2) Q(p_1), G(k_2,\varepsilon_2,a) \vert J^\mu (0) \vert 0 \rangle \vert_{Tree\ level}
  +\cdots,
\end{equation}
where we only give those contributions explicitly, which are from the color-octet $Q\bar Q$ pair
with the quantum numbers $^1S_0$ and $^3P_J$ with $J=0,2$. The contributions
from states different than those above are indicated by $\cdots$. They are not important
for our purpose, but they are important for the complete cancellation of soft divergences in the KLN
theorem. We will discuss this later. The contribution from $R^\mu_{gg,Q,s}$ to $T^{\mu\nu}$ can be calculated
easily, we obtain:
\begin{equation}
T^{\mu\nu}_{1,Q,s}(P,q_\gamma,Q\bar Q) = \frac{\alpha_s}{2\pi} N_c \left ( \frac{2}{\epsilon_{I}} \right )
\left [ 1+\frac{1}{3} v^2 +\frac{1}{3} v'^2 +{\mathcal O}(v^4) \right ]
T_0^{\mu\nu} (P,q_\gamma, Q\bar Q)+\cdots
\end{equation}
where we have only collected the contributions from relevant $Q\bar Q$ states. Contributions
from other states and higher orders of $v$ are represented by $\cdots$.
\par
The amplitude from the gluon splitting is given by:
\begin{eqnarray}
R^\mu_{gg,g} & =&   eQ g_s^2 f^{aa_1a_2}  \bar u(p_1) T^a \left [ \gamma^\nu \frac{\gamma\cdot(p_1+k)+m}{(p_1+k)^2 -m^2}
\gamma^\mu + \gamma^\mu \frac{\gamma\cdot(-p_2 -k)+m}{(p_2+k)^2 -m^2} \gamma^\nu \right ] v(p_2)
\nonumber\\
  && \left [ (k+k_1)\cdot\varepsilon_2^* \varepsilon^*_{1\nu} +(-k_1+k_2)_\nu \varepsilon_1^* \cdot \varepsilon_2^*
        +(-k_2-k)\cdot\varepsilon_1^*  \varepsilon_{2\nu}^* \right ] \frac{1}{(k_1+k_2)^2},
\end{eqnarray}
with $k=k_1 +k_2$. The contribution from $R^\mu_{gg,g}$ to $T^{\mu\nu}$ is more difficult to evaluate
than that from $R^\mu_{gg,Q}$, because there is a overlap between the collinear- and soft region
of gluon momenta. However there is a standard method, called phase-space slicing method, discussed
in detail in \cite{Arnd,GG}. We will use the method.
If the gluon with $k_1$ is soft, the amplitude can be approximated by:
\begin{eqnarray}
R^\mu_{gg,g,s} & = &   eQ g_s^2 f^{aa_1a_2}
\bar u(p_1)T^a \left [ \gamma^\nu \frac{\gamma\cdot(p_1+k_2)+m}{(p_1+k_2)^2 -m^2}
\gamma^\mu + \gamma^\mu \frac{\gamma\cdot(-p_2 -k_2)+m}{(p_2+k_2)^2 -m^2} \gamma^\nu \right ] v(p_2)
\nonumber\\
       && \cdot  \frac { (-2 k_2 \cdot\varepsilon_1^* )   \varepsilon_{2\nu}^* }{(k_1+k_2)^2}.
\end{eqnarray}
This amplitude $R^\mu_{gg,g,s}$ interfered with the $R^\mu_{gg,Q,s}$ will give contributions
with the soft divergences which are exactly those in the contributions from the interference
between $R^\mu_{gg,Q}$ and $R^\mu_{gg,g}$. By the method mentioned above we have the contribution
from the interference:
\begin{eqnarray}
T^{\mu\nu}_{1,int.}(P,q_\gamma,Q\bar Q) = \frac{\alpha_s}{2\pi} N_c \left \{
 \left ( \frac{2}{\epsilon_I}\right )^2 - \frac{2}{\epsilon_I} \left [
   \gamma + \ln\frac{s^2_{min}}{\pi m^2 \mu^2}\right ]\right \}
T_0^{\mu\nu} (P,q_\gamma, Q\bar Q),
\end{eqnarray}
where $s_{min}$ is a cut-off in the phase-space slicing method. Our final result will
not depend on it.
The contributions only from the gluon splitting into two gluons and a light quark pair can be calculated
directly with the phase-space slicing method. They contain only collinear singularities. The result
is:
\begin{eqnarray}
T^{\mu\nu}_{1, col}(P,q_\gamma,Q\bar Q) = \frac{\alpha_s}{2\pi} \left ( \frac{2}{\epsilon_I}\right )
  \left \{ N_c \left [ - L_n +\ln\frac{s^2_{min}}{\pi m^2 \mu^2}
     +\frac{11}{6}\right ]
      -\frac{N_f}{3} \right \}
T_0^{\mu\nu} (P,q_\gamma, Q\bar Q),
\end{eqnarray}
where the same dependence on the cut-off $s_{min}$ appears and it will be cancelled by that
in $T^{\mu\nu}_{1,int.}$.
Putting everything together we obtain the infrared divergent part of the real correction as:
\begin{eqnarray}
 T^{\mu\nu}_{1,\ real}(P,q_\gamma,Q\bar Q) &=&  \frac{\alpha_s}{2\pi} \left \{ N_c \left [
 \left ( \frac{2}{\epsilon_I}\right )^2 - \frac{2}{\epsilon_I} \left (
   \gamma -1  + L_n \right ) \right ] \right \}
   T_0^{\mu\nu} (P,q_\gamma, Q\bar Q)
\nonumber\\
  &&  +\frac{\alpha_s}{2\pi} \frac{2}{\epsilon_I}\left \{
    N_c\left [ \frac{11}{6}+\frac{1}{3} v^2+ \frac{1}{3} v'^2 \right ] -\frac{N_f}{3} \right \}
T_0^{\mu\nu} (P,q_\gamma, Q\bar Q) +\cdots,
\end{eqnarray}
Finally, we obtain the total soft-divergent part of $T^{\mu\nu}$ at one loop:
\begin{eqnarray}
T_{1}^{\mu\nu}(P,q_\gamma, Q\bar Q)
  & =& \frac{\alpha_s }{2\pi}\left [(v^2+v'^2) \left (\frac{2}{\epsilon_I}\right )\frac{N_c^2-2}{3N_c}
                 - \frac{1}{N_c} \frac{\pi^2}{2v} \right ] T_0^{\mu\nu} (P,q_\gamma, Q\bar Q)  +\cdots,
\end{eqnarray}
where we only give the relevant part in detail. It should be noted that
the $[\cdots]$ of the part does
not contain ${\bf v}\cdot {\bf v'}$ from our calculation at the orders considered here.
The $\cdots$ denote contributions from other states of the $Q\bar Q$ pair, which can not be produced
at tree-level. These contributions are from
$R^\mu_{gg,Q}$, represented by $\cdots$ in Eq.(18), they contain terms like ${\bf v}\cdot {\bf v'}$
because the $Q\bar Q$ pair at one-loop order can be in a color-singlet $P$-wave states.
\par
Before we turn to our result of NRQCD matrix elements to finalize NRQCD factorization, we
briefly discuss here the cancellation of soft divergences in KLN theorem.
It should be noted that in $T^{\mu\nu}$ defined in Eq.(5) the heavy quark pair can be
in an arbitrary state which is allowed to be produced at a given order of $\alpha_s$,
i.e., we do not sum over all possible states of the heavy quark pair.
If we sum all these states and take ${\bf v} ={\bf v'}$, the sum should
not contain any soft divergence, as stated by KLN theorem. The sum contains not only those terms
explicitly given in Eq.(24), but also those represented by $\cdots$, i.e., the contributions
from those states which can not be produced at tree level, these states are produced through
emission of two gluons by heavy quarks, whose contributions are represented in Eq.(18) by $\cdots$.
Taking these contributions into account we have the soft-divergent part of the sum:
\begin{eqnarray}
\sum T_{1}^{\mu\nu}(P,q_\gamma, Q\bar Q){\Big\vert}_{{\bf v} ={\bf v'}}
   &=& \frac{\alpha_s }{\pi}v^2 \left (\frac{2}{\epsilon_I}\right )\frac{N_c^2-2}{3N_c}
 \sum T_0^{\mu\nu} (P,q_\gamma, Q\bar Q){\Big\vert}_{{\bf v} ={\bf v'}}
\nonumber\\
   && -\frac{\alpha_s }{\pi}v^2 \left (\frac{2}{\epsilon_I}\right )\frac{N_c^2-2}{3N_c}
 \sum T_0^{\mu\nu} (P,q_\gamma, Q\bar Q){\Big\vert}_{{\bf v} ={\bf v'}} +\cdots ,
\end{eqnarray}
where the second term comes from the contributions of those states which are not given explicitly in Eq.(18).
By summing of these states and some manipulation the sum can be written into a compact form
as given in the above.
We see clearly that the sum is free from any soft divergence as KLN theorem states.
\par
%%%%%%%%%%%%%%%% Inset Fig. 1 here %%%%%%%%%%%%%%%%%%%%%%%%%%%%%%

\begin{figure}[hbt]
\begin{center}
\includegraphics[width=9cm]{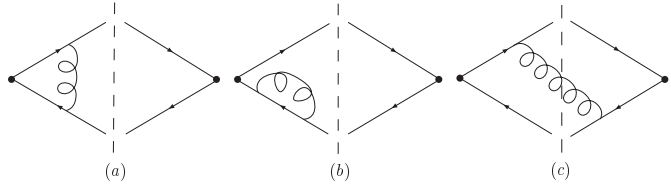}
\end{center}
\caption{ Examples of Feynman diagrams for (a) vertex correction, (b) wave-function renormalization and (c)
gluon-exchange correction, the broken line is the cut.}
\label{Feynman-dg2}
\end{figure}
%\end{center}
\par
The one-loop correction of those NRQCD matrix elements in Eq.(8) and Eq.(12) can be divided into two parts as
a virtual part and a real part.
The virtual part consists of corrections from vertex and wave-function renormailization.
Examples of Feynman diagrams at one-loop level for the real part, vertex- and wave-function
renormailization correction are given in Fig.3.
up to the orders of $v$ we consider.
We calculate these corrections in Feynman gauge.
The one loop correction can be written as:
\begin{equation}
\langle 0 \vert O_n^{Q\bar Q} \vert 0 \rangle_1
= \langle 0 \vert O_n^{Q\bar Q} \vert 0 \rangle_0 \left [
 -\frac{\pi \alpha_s}{4v} \cdot \frac{1}{N_c}
   (1+ {\mathcal O}(v^2))
    + {\mathcal O}(v^2) \right ]
\end{equation}
where $O_n^{Q\bar Q}$ is any of $O_8(^1S_0)$, $P_8(^1S_0)$, $O(^3P_J)$ with $J=0,1,2$,
the matrix element with the subscriber $0(1)$
is the tree-level(one-loop) result of the operator.
The first term represents the Coulomb singularity,
the second term starts at the relative order of $v^2$, combined with
$\langle 0 \vert O_n^{Q\bar Q} \vert 0 \rangle_0$ they should be written in the form
as matrix elements of operators.
Using this result we can clearly see that
the Coulomb singularity in Eq.(24) is reproduced. Hence, one can conclude
that all perturbative coefficients
$F's$ in Eq.(3) are free from the Coulomb singularity and also from I.R. divergences.
\par
To assess if $G^{\mu\nu}(^1S_0)$ contains any I.R. divergence, we need to take
the operator mixing between $O_8(^1S_0)$ and others into account which is contained
in the second term in Eq.(26). At one loop and order $v^2$, the operator
can be mixed with the color-singlet operator with quantum numbers $^1P_1$ and
the color-octet operator $P_8(^1 S_0, Q\bar Q)$. The later is relevant for our case.
In Feynman gauge we have
\begin{eqnarray}
\langle 0 \vert O_8(^1 S_0, Q\bar Q)\vert 0\rangle
   &=& -\frac{\alpha_s}{2\pi} \frac{1}{N_c}  \left(\frac{2}{\epsilon_I} \right )
    \left ( \langle 0 \vert O_8(^1 S_0, Q\bar Q)\vert 0\rangle_0
    + \frac{2}{3 m^2} \langle 0 \vert P_8(^1 S_0, Q\bar Q)\vert 0\rangle_0 +{\mathcal O} (v^4) \right )
\nonumber\\
     &&   -\frac{\alpha_s}{2\pi}\frac{N_c^2-1}{N_c} \left (\frac{2}{\epsilon_I}\right )
      \langle 0 \vert O_8(^1 S_0, Q\bar Q)\vert 0\rangle_0
\nonumber\\
   && +\frac{\alpha_s}{2\pi} N_c \left (\frac{2}{\epsilon_I} \right )
    \left (  \langle 0 \vert O_8(^1 S_0, Q\bar Q)\vert 0\rangle_0
     +\frac{1}{3 m^2}  \langle 0 \vert P_8(^1 S_0, Q\bar Q)\vert 0\rangle_0 \right )
\nonumber\\
   && + \cdots,
\nonumber\\
\langle 0 \vert P_8(^1 S_0, Q\bar Q)\vert 0\rangle_0 &=&
m^2(v^2 + v'^2) \langle 0 \vert O_8(^1 S_0, Q\bar Q)\vert 0\rangle_0,
\end{eqnarray}
where contributions at orders higher than $v^2$ or from mixing of other irrelevant
operators and those  with the Coulomb singularity are represented
by $\cdots$. In the above
the first line comes from the vertex correction, the second from the external quark legs and
the third comes from the real gluon exchange.
With this result, one can see clearly how the soft divergences are absorbed into the matrix element
at leading order of $v$ on a diagram-by-diagram basis.
At order of $v^0$ all I.R. divergences are cancelled.
At the next-to-leading order of $v$, the net divergence
in the above equation exactly matches that in $T_1^{\mu\nu}$ and the matching is also
on a diagram-by-diagram basis.
\par
\par
%%%%%%%%%%%%%%%% Inset Fig. 1 here %%%%%%%%%%%%%%%%%%%%%%%%%%%%%%

\begin{figure}[hbt]
\begin{center}
\includegraphics[width=7cm]{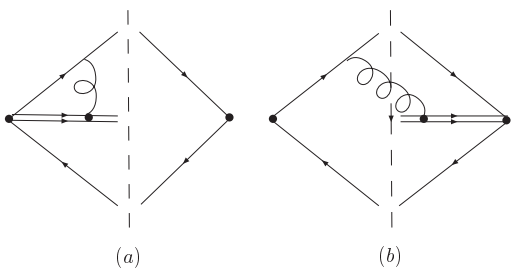}
\end{center}
\caption{ Types of Feynman diagrams for contributions with the gauge link.
The double line stands for the gauge link.
(a): Virtual correction. (b): Real correction.}
\label{Feynman-dg2}
\end{figure}
%\end{center}
\par
Our results clearly show that without the modification of
relevant NRQCD matrix elements in our case the NRQCD factorization holds at order of $v^2$ and
the relativistic correction can also be factorized. Adding a gauge link
in the those color-octet NRQCD matrix elements, as suggested in \cite{Qiu},
the modified matrix elements will receive extra contributions.
These contributions relevant for our study are given by types of Feynman diagrams in Fig.4.
We use Feynman gauge to study these contributions. In other gauges, like Coulomb gauge,
a gluon can also be exchanged between gauge links.
By an infrared power counting each contribution from Fig.4. is with infrared divergences.
The virtual contribution from the type(a) of diagrams in Fig.4, after integrating of the energy of the virtual gluon
and neglecting terms which generate power divergences, is proportional to
the integral:
\begin{eqnarray}
\int \frac{ d^3 q}{(2\pi)^3} \frac{1}{2\vert {\bf q}\vert} \frac{1}{\vert {\bf q}\vert + q^3}
 \left [ \frac{ 1 + v^3} { -\vert {\bf q}\vert +{\bf v} \cdot {\bf q}}
  +\frac{ 1 - v^3} { -\vert {\bf q}\vert  -{\bf v} \cdot {\bf q}} \right ].
\end{eqnarray}
In the above the factor $1/(\vert {\bf q}\vert + q^3)$ comes from the eikonal propagator
of the gauge link, which is determined by the moving direction of the quarkonium. We take
the direction in the $z$-direction. It is clearly that the integral is infrared-divergent.
But the real contribution from the type(b) of diagrams in Fig.4 is also proportional
to this integral and the proportional coefficient has different sign than that
of the virtual contribution. The total contribution from the gauge link
to the color-octet matrix elements is zero at one loop and at any order of $v$.
Therefore, at one-loop level, the suggested gauge link will not lead to a violation
of the universality of color-octet matrix elements and it will not affect all existing one-loop
results.
\par
To summarize:  We have examined in detail the NRQCD factorization in inclusive production
of a quarkonium through $e^+e^-$-annihilation. Our results show that the factorization
can be made for production of a $Q\bar Q$ pair in color octet with the quantum number
$^1S_0$ and $^3P_{0,2}$ and also for relativistic correction to the $S$-wave state.
The modification of color-octet NRQCD matrix elements with the suggested gauge link\cite{Qiu}
will not affect the NRQCD factorization at one loop in cases studied before and the case studied here,
because the gauge link does not lead to nonzero contributions to color-octet NRQCD matrix elements
at one loop. Because of this the universality of these matrix elements
holds at one loop.

\vskip20pt
\noindent
{\bf Acknowledgements}
\par
The authors would like to thank Prof. J.W. Qiu
for communications about the recent work\cite{Qiu}
and Prof. G. Bodwin for intensive discussions, which greatly help
to understand the problem studied here.
The warm hospitality  of the Taipei Summer Institute
of NCTS/TPE at the Physics Department of NTU, where the first draft of
the paper is completed, is acknowledged.
This work is supported by National Nature
Science Foundation of P. R. China.

\par

\end{document}